\documentclass[preprint2]{aastex63}

\received{11-22-2019}
\revised{2-11-2020}
\accepted{3-20-2020}
\submitjournal{ApJ}

\usepackage[version=4]{mhchem}
\defcitealias{nationalaeronauticsandspaceadministrationExoplanetExplorationPlanets2019}{https://exoplanets.nasa.gov}
\graphicspath{ }
\usepackage{xcolor}
\definecolor{darkolivegreen}{rgb}{0.33, 0.42, 0.18}
\usepackage{natbib}
\usepackage{amsmath}
\usepackage[normalem]{ulem}
\usepackage{lineno}

\shorttitle{Detectability of Life on Pelagic Planets and Water Worlds}
\shortauthors{Glaser et al.}

\begin{document}

\title{Detectability of Life Using Oxygen on Pelagic Planets and Water Worlds}

\correspondingauthor{Hilairy Ellen Hartnett}
\email{h.hartnett@asu.edu}
\affiliation{School of Molecular Sciences, Arizona State University}
\affiliation{School of Earth and Space Exploration, Arizona State University}

\correspondingauthor{Steven J Desch}
\email{steve.desch@asu.edu}
\affiliation{School of Earth and Space Exploration, Arizona State University}

\author[0000-0002-1150-4605]{Donald M Glaser}
\affiliation{School of Molecular Sciences, Arizona State University}

\author[0000-0003-0736-7844]{Hilairy Ellen Hartnett}
\affiliation{School of Molecular Sciences, Arizona State University}
\affiliation{School of Earth and Space Exploration, Arizona State University}

\author[0000-0002-1571-0836]{Steven J Desch}
\affiliation{School of Earth and Space Exploration, Arizona State University}

\author[0000-0001-8991-3110]{Cayman T Unterborn}
\affiliation{School of Earth and Space Exploration, Arizona State University}

\author{Ariel Anbar}
\affiliation{School of Molecular Sciences, Arizona State University}
\affiliation{School of Earth and Space Exploration, Arizona State University}

\author{Steffen Buessecker}
\affiliation{School of Life Sciences, Arizona State University}

\author{Theresa Fisher}
\affiliation{School of Earth and Space Exploration, Arizona State University}

\author{Steven Glaser}
\affiliation{School of Earth and Space Exploration, Arizona State University}

\author{Stephen R Kane}
\affiliation{Department of Earth and Planetary Sciences, University of California, Riverside}

\author{Carey M Lisse}
\affiliation{Applied Physics Laboratory, Johns Hopkins University}

\author[0000-0002-4579-3628]{Camerian Millsaps}
\affiliation{School of Earth and Space Exploration, Arizona State University}

\author{Susanne Neuer}
\affiliation{School of Life Sciences, Arizona State University}

\author[0000-0002-1180-996X]{Joseph G O’Rourke}
\affiliation{School of Earth and Space Exploration, Arizona State University}

\author{Nuno Santos}
\affiliation{Instituto de Astrof\'{i}sica e Ci\^{e}ncias do Espa\c{c}o, Universidade do Porto}

\author{Sara Imari Walker}
\affiliation{School of Earth and Space Exploration, Arizona State University}

\author{Mikhail Zolotov}
\affiliation{School of Earth and Space Exploration, Arizona State University}



\begin{abstract}

The search for life on exoplanets is one of the grand scientific challenges of our time. The strategy to date has been to find (e.g., through transit surveys like Kepler) Earth-like exoplanets in their stars’ habitable zone, then use transmission spectroscopy to measure biosignature gases, especially oxygen, in the planets’ atmospheres (e.g., using JWST, the James Webb Space Telescope). Already there are more such planets than can be observed by JWST, and missions like the Transiting Exoplanet Survey Satellite and others will find more. A better understanding of the geochemical cycles relevant to biosignature gases is needed, to prioritize targets for costly follow-up observations and to help design future missions. We define a Detectability Index to quantify the likelihood that a biosignature gas could be assigned a biological vs. non-biological origin. We apply this index to the case of oxygen gas, O\textsubscript{2}, on Earth-like planets with varying water contents. We demonstrate that on Earth-like exoplanets with ~0.2 weight percent (wt\%) water (i.e., no exposed continents) a reduced flux of bioessential phosphorus limits the export of photosynthetically produced atmospheric O\textsubscript{2} to levels indistinguishable from geophysical production by photolysis of water plus hydrogen escape. Higher water contents $>$1wt\% that lead to high-pressure ice mantles further slow phosphorus cycling. Paradoxically, the maximum water content allowing use of O\textsubscript{2} as a biosignature, ~0.2wt\%, is consistent with no water based on mass and radius. Thus, the utility of an O\textsubscript{2} biosignature likely requires the direct detection of both water and land on a planet.

\end{abstract}



\section{Introduction} \label{sec:intro}
\subsection{The Search for Life on Exoplanets}
One of the grand challenges of our time is to find the signs of life on planets around other stars. Whether there is life elsewhere in the universe is a question that people have pondered for millennia, but only in the last few decades have the tools been acquired to observe exoplanets and potentially answer the question. The identification of the signs of life on an exoplanet where no life existed, or the failure to detect it on an exoplanet bearing life are all-too-possible outcomes, and would be equally tragic. 

The current strategy for finding life on exoplanets requires two levels of astronomical observation: exoplanet classification, and atmospheric characterization. First, radial velocity, transit, and/or transit timing variation (TTV) methods can identify the mass, planet radius, and orbital radius that together can be used to classify an exoplanet as “Earth-like” and whether it is in its star’s habitable zone. Second, visible and infrared transmission spectroscopy or reflectance spectroscopy can characterize atmospheric gases such as ozone, diatomic oxygen, carbon dioxide, water vapor, and possibly methane \citep{demingDiscoveryCharacterizationTransiting2009, domagal-goldmanAstrobiologyPrimerV22016, kalteneggerHowCharacterizeHabitable2017, kalteneggerTransitsEarthLikePlanets2009, lovisAtmosphericCharacterizationProxima2017, meadowsReflectionsO2Biosignature2017, meadowsExoplanetBiosignaturesUnderstanding2018, schwietermanExoplanetBiosignaturesReview2018}. The presence of oxygen (i.e., O\textsubscript{2} and O\textsubscript{3}) in the atmosphere of a rocky planet is widely considered a robust biosignature \citep{meadowsExoplanetBiosignaturesUnderstanding2018}.

The pace of planet detection and characterization has increased exponentially over the past decade and now the exoplanet community has an abundance of potential astrobiologically-relevant targets. The Kepler mission identified over 2600 exoplanets. There already may be (and certainly will be) more “Earth-sized” (i.e., radius, R $<$ 1.5 R\textsubscript{Earth}), rocky exoplanets in their stars’ habitable zones than can be analyzed with follow-up observations \citep{fultonCaliforniaKeplerSurvey2017}. The new Transiting Exoplanet Survey Satellite (TESS) is currently identifying exoplanets, and future missions like the PLAnetary Transits and Oscillations of stars (PLATO) will find thousands more planets, potentially adding dozens more to the list of exoplanets in their stars’ habitable zones, further expanding the backlog of “Earth-like” exoplanets for follow-up observations \citep{rauerPLATOMissionPLATO2016, sullivanTransitingExoplanetSurvey2015}. These follow-up studies are time-consuming, and there will not be sufficient telescope time available to probe the atmospheres of all of these habitable rocky exoplanets \citep{morleyObservingAtmospheresKnown2017}. To emphasize this point, the James Webb Space Telescope (JWST) is estimated to be able to characterize the atmospheres of only 5 exoplanets over its mission lifetime, creating the need to maximize the chances of a ‘true positive’ detection \citep{brownNewCompletenessMethods2010}. At present, the exoplanet community is beginning to develop systematic guiding principles to help select targets and prioritize exoplanets for follow-up observations \citep{meadowsExoplanetBiosignaturesUnderstanding2018, deschProcedureObservingRocky2018, nationalacademiesExoplanetScienceStrategy2018}. The search criterion that has served the community well until now—habitability, as defined by orbital parameters allowing the possibility of sustaining liquid water on the surface—is no longer sufficiently restrictive.
\subsection{Moving Beyond 'Habitability' in the Search for Life}
We advocate use of a more restrictive criterion for exoplanet target selection: detectability of life. Of course, to detect life, life must be able to exist on a planet, so we cannot discard the concept of habitability. However, habitability as a criterion is not especially selective and is no longer sufficient: we have or will soon discover more than enough planets that are conservatively “habitable.”

Oxygen may be the best biosignature for astrobiology due to the relative ease with which it can be measured \citep{catlingWhyO2Required2005}. On Earth, atmospheric oxygen exists because of life. But oxygen will not indicate life on all planets \citep{meadowsReflectionsO2Biosignature2017, meadowsExoplanetBiosignaturesUnderstanding2018}. A number of “false positive” scenarios (e.g., for Venus and Mars, or for water-rich planets around M dwarfs) have been identified that could give rise to abundant oxygen in the atmosphere of a planet by purely non-biological processes \cite[e.g.,][]{domagal-goldmanAbioticOzoneOxygen2014, lugerExtremeWaterLoss2015, meadowsReflectionsO2Biosignature2017}. Disentangling the geological and biological oxygen signals requires detailed knowledge of a planet’s geological processes and geochemical cycles. There are many exoplanets where life may very well exist but would not be detectable. What the community needs is a continuous-scale criterion for selecting those exoplanets for which life, if it exists, would be detectable and discernible against the non-biological, geochemical background of the planet.
\section{Quantifying Detectability: The Detectability Index}
Habitability as a criterion to inform future time-intensive spectroscopy of exoplanet atmospheres is insufficient as it is a discrete, binary output: ’habitable’ or ’not-habitable’. Detectability, however, can be conceptualized and quantified on a continuous scale, such that given any number of habitable exoplanets we can calculate a detectability index of a biosignature molecule that considers the geochemical context. This allows astronomers and astrobiologists to maximize the likelihood of an unambiguous signal of life. To be quantitative in our assessments of whether a biological signature would be discernible above the geochemical background of a planet, we define a Detectability Index (DI) unique for every putative biosignature molecule in the atmosphere of a planet:
\begin{equation}
	\begin{split}
		\begin{aligned}[left]
			\label{DI:General}
			&{\scriptstyle DI({\rm gas, planet}) = }\\
			&{\scriptstyle \log_{10}(\dfrac{\scriptstyle Maximum\;net\;biological\;gas\;production\;rate}{\scriptstyle Gross\;geological\;gas\;production\;rate})}.
		\end{aligned}
	\end{split}
\end{equation}
This equation uses estimated or modeled rates of biosignature molecule creation and destruction by biological and non-biological sources to quantify the relative magnitude of a biological signal. DI is expressed as the logarithm of a potential biosignature gas, because the rates can vary by several orders of magnitude. To do so, it is important to review how life would be detected on an exoplanet using a biosignature gas; here, we use O\textsubscript{2} as an informative example.

It is important to note that the presence of a gas does not by itself indicate life; life is inferred only if the presence of a gas demands production of the gas at rates only attributable to life. Transmission spectroscopy can be used to determine the abundance of O\textsubscript{2} in a planet’s atmosphere. On Earth atmospheric oxygen is consumed by reactions with reduced species outgassed from Earth’s interior, or with surface minerals \citep{catlingGreatOxidationEvent2014}. Assuming similar processes operate on an exoplanet, the rate of destruction of atmospheric O\textsubscript{2} could be inferred. Assuming steady state, this would equal the rate of oxygen production by all sources, non-biological plus biological. Thus, the total production rate of O\textsubscript{2} might be inferred. In order for a production rate to be attributable to life, the non-biological production rate must be known \textit{a priori} to be smaller than at least some likely values of the biological production rate. Our Detectability Index distinguishes those planets for which a gas, in this case oxygen, could be a definite biosignature, from those for which oxygen is not a clear biosignature. In this context, “not a clear biosignature” does not mean oxygen is not produced by life; it means oxygen is not \textit{definitively} produced by life. In other words, it is necessary to disentangle the nonbiological and biological sources and sinks of a biosignature gas. Here we describe two novel ideas in astronomy. First, we show necessity of the detectability index to identify potentially fruitful astrobiological exoplanet targets. Second, we demonstrate the utility of the detectability index by assessing the potential detectability of life on Earth-sized planets with increased surface water.
\section{Detectability Index Framework Benchmark: Earth}
Here we calculate a Detectability Index for O\textsubscript{2} on Earth, where a balance of biological and geological sources and sinks maintains the steady state atmospheric oxygen concentration ([O\textsubscript{2}]), i.e., net oxygen production:
\begin{equation}
	\begin{split}
		\label{O2:net}
		&{\scriptstyle Net\;{\rm O_{2}}\;production = }\\
		&{ \scriptstyle [Source_{\rm {geo}} - Sink_{\rm {geo}}] + [Source_{\rm {bio}} - Sink_{\rm {bio}}]  = 0,}
	\end{split}
\end{equation}
where Source\textsubscript{geo} is photolysis, Sink\textsubscript{geo} is oxidation of reduced species, Source\textsubscript{bio} is oxygenic photosynthesis, Sink\textsubscript{bio} is aerobic respiration, [Source\textsubscript{bio} – Sink\textsubscript{bio}] = net biological O\textsubscript{2} production, and [Source\textsubscript{geo} – Sink\textsubscript{geo}] = net geological O\textsubscript{2} production. On Earth, oxygen is considered a valid biosignature because it is overwhelmingly produced by life (we refer the reader to Catling’s \citeyearpar{catlingGreatOxidationEvent2014} comprehensive review of oxygen on Earth). Oxygenic photosynthesis on Earth today produces a gross flux (Source\textsubscript{bio}) of $\sim$10\textsuperscript{4} Tmol O\textsubscript{2} yr\textsuperscript{-1} (1 Tmol yr\textsuperscript{-1} = 10\textsuperscript{12} moles per year) through the following reaction:
\begin{equation}
\label{eq:photo}
{\scriptstyle \ce{CO2 + H2O ->[h$\nu$] \text{``}CH2O\text{''} + O\textsubscript{2},}}
\end{equation}
where “CH\textsubscript{2}O” represents a simplified carbohydrate to balance the chemical reaction. More generally, “CH\textsubscript{2}O” is shorthand for the organic carbon produced by organisms that also carries nitrogen, phosphorus, sulfur, and trace elements in broadly universal stoichiometric amounts. Organic carbon and O\textsubscript{2} are produced in a 1:1 stoichiometric ratio. Aerobic respiration (i.e., organic carbon oxidation, Sink\textsubscript{bio}) is the reverse reaction and consumes a nearly equivalent $\sim$10\textsuperscript{4} Tmol O\textsubscript{2} yr\textsuperscript{-1}. Some organic carbon is oxidized via anaerobic respiration, but the reduced species produced (e.g., S\textsuperscript{2-}, or Fe\textsuperscript{2+}) are assumed to be re-oxidized by O\textsubscript{2}, so the moles of oxygen are not double counted \citep{hartnettRoleStrongOxygendeficient2003}. The imbalance between production and respiration is ultimately the biological source of atmospheric oxygen. When one mole of organic carbon is buried in marine sediments and escapes degradation, the corresponding one mole of O\textsubscript{2} is ‘stranded’ in the atmosphere. On Earth today, the organic carbon burial rate corresponds to a net biological production of $\sim$20 Tmol O\textsubscript{2} yr\textsuperscript{-1}.

The non-biological O\textsubscript{2} production rate (by photolysis of H\textsubscript{2}O and subsequent hydrogen escape, i.e., Source\textsubscript{geo}) is three orders of magnitude smaller, only 0.02 Tmol O\textsubscript{2} yr\textsuperscript{-1}. Oxygen is consumed (Sink\textsubscript{geo}) by reduced iron and reduced sulfur minerals or dissolved ions (e.g., Fe\textsuperscript{2+} and HS\textsuperscript{-}), derived from rocks (the so-called “rock buffer”). Oxygen also reacts with reduced gas species supplied through volcanism or serpentinization (e.g., H\textsubscript{2}, H\textsubscript{2}S, CO, and CH\textsubscript{4}). These yield a lifetime of an O\textsubscript{2} molecule against destruction. To first order, the total destruction rate is proportional to the number of moles of O\textsubscript{2} in the atmosphere, N\textsubscript{O\textsubscript{2}}, divided by the lifetime ($\tau$); and the rate of change of atmospheric oxygen in steady state is the production rate minus the destruction rate, which is zero in steady state.

\begin{equation}
	\begin{split}
		\label{O2:rateSteadys}
		&{\scriptstyle \dfrac{\scriptstyle dN_{\rm {O_{2}}}}{\scriptstyle dt} = }\\
		&{\scriptstyle Source_{\rm {geo}} + (Source_{\rm {bio}} - Sink_{\rm {bio}}) {\scriptstyle -  }\dfrac{\scriptstyle N_{\rm {\scriptstyle O_{2}}}}{\tau} = 0}
	\end{split}
\end{equation}
There are exceptions to this — for example, oxygen is not the limiting species in reaction with massive sulfides at ridge axes \citep{sleepDioxygenGeologicalTime2005} — but destruction of O\textsubscript{2} is dominated by reaction with reduced volcanic gases and reduced minerals \citep{catlingGreatOxidationEvent2014}. Estimation of the lifetime is necessary to convert the observed abundance of O\textsubscript{2} in the atmosphere into a destruction rate, and therefore a total production rate. The rate of consumption of O\textsubscript{2} via respiration (Sink\textsubscript{bio}) and reduced species oxidation (Sink\textsubscript{geo}) proceeds on Earth today at approximately 20 Tmol O\textsubscript{2} yr\textsuperscript{-1}, balancing the biological net production of O\textsubscript{2}, and creating an instantaneous steady state 20\% O\textsubscript{2} mixing ratio in the atmosphere. 
It follows that DI for oxygen on Earth is:

\begin{equation}
\label{DI:Earth}
	\begin{split}
		\begin{aligned}
			&{\scriptstyle DI({\rm O_{2}, Earth}) =}\\
			&{\scriptstyle \log_{10}(\dfrac{\scriptstyle Net\;biological\;{\rm O_{2}}\;production}{\scriptstyle Geological\;{\rm O_{2}}\;production}) = }\\
			&{\scriptstyle \log_{10}(\dfrac{\scriptstyle (Source_{\rm {bio}} - Sink_{\rm {bio}})}{\scriptstyle Source_{\rm {geo}}})}\\[10pt]
			&{\scriptstyle DI({\rm O_{2}, Earth}) = }\\
			&{\scriptstyle \log_{10}(\dfrac{\scriptstyle 20\;{\rm Tmol\;yr^{-1}}}{\scriptstyle 0.02\;{\rm Tmol\;yr^{-1}}}) = 3.}
		\end{aligned}
	\end{split}
\end{equation}

In this example the (net) biological production rate of O\textsubscript{2} exceeds the geological production rate by three orders of magnitude. Because we can estimate geochemical rates to reasonable precision and we know that DI(O\textsubscript{2}, Earth) is a large positive number, we recognize oxygen as a valid biosignature on Earth, and thus, possibly on similar exoplanets. A DI $\leq$ 0 on an exoplanet would indicate that net biological production rates do not exceed geological rates; on such a planet we could not conclude with confidence that any of the total production rate of the gas (here, O\textsubscript{2}) is attributable to biological processes.

Our ability to infer life requires knowledge of geochemical cycles. While the geochemical cycles on Earth are constrained, albeit to only tens of percent, it is unlikely that we can infer the geochemical cycles on even the best resolved exoplanet with anywhere near the precision we have for Earth. However, geochemical cycles can be approximated, allowing us to identify exoplanets for which oxygen will not be a biosignature. This will allow us to estimate a Detectability Index and systematically prioritize other exoplanets for limited follow-up observations. Planets for which geological production rates of oxygen likely dominate over the possible biological production rates of oxygen will have DI(O\textsubscript{2}, exoplanet) $<$ 0. These planets should be de-prioritized for observations searching for signs of photosynthetic life because any oxygen in the planet’s atmosphere always could be solely attributed to geological processes. Planets for which estimated geological and biological rates are approximately equal (DI $\sim$0–1), while better, also would be difficult to assess because one still could not definitively attribute, within uncertainties, a biological origin to the atmospheric oxygen. Atmospheric oxygen could only be ambiguously attributed to life. Only those exoplanets with the largest, positive values of DI(O\textsubscript{2}, exoplanet) should be prioritized for future observations in our search for life.

This approach goes far beyond questions of habitability and whether liquid water can exist on the surface. It even goes beyond assumptions that O\textsubscript{2} is always a robust biosignature gas except perhaps for a limited number of scenarios where it is a false positive. That leads to binary decisions about exoplanets: oxygen is not considered a biosignature on some exoplanets, but is considered as good a biosignature as on Earth for all other planets. Our approach recognizes that the reliability of a biosignature is on a sliding scale, which helps us to prioritize exoplanets for follow-up observations. This approach demands that we strive for an integrated understanding of the geochemical cycles of an exoplanet, using as much compositional and geophysical information as possible in order to constrain the expected range of geological rates. Below we show that currently available measurements (e.g., planet radius), modeled properties (e.g., bulk water content and surface temperature), and approximated geochemical cycles can be used to quantify the detectability of oxygen as a biosignature on planets with significant amounts of water.
\subsection{Detectability Index of Oxygen on Earth}
We first consider the detectability index in the context of Earth. We begin by reviewing how geochemical cycles control the gross production and accumulation of oxygen on this planet. Earth’s biosphere maintains a net O\textsubscript{2} production rate that is 1000 times the gross geological oxygen production rate, creating a strong oxygen biosignature (see eq. \ref{DI:Earth}). This is a large ratio, but is surprisingly dependent on many factors: not just the presence of liquid water and habitable temperatures, but also the fluxes of nutrients and trace metals. On Earth today, these are provided and maintained through geochemical cycling of the surface environment. How to translate the abundance of oxygen in an atmosphere to a production rate demands quantifying the destruction rate of oxygen, which is surprisingly dependent on other factors. For example, the geological conditions of a planet may be sufficiently reducing as to consume most or all of the biological oxygen production (i.e., Sink\textsubscript{geo} $>>$ (Source\textsubscript{bio} + Source\textsubscript{geo})), as may have been the case during the Archean eon of Earth’s history. 

To demonstrate the effectiveness of the detectability index in helping to prioritize observational targets based the plausible robustness of their candidate biosignatures, we start with the example of Earth’s oxygen. Here we describe three categories of geological processes that are necessary to detect atmospheric oxygen. First, the surface must be habitable; by that we mean that the planet’s climate must be regulated such that liquid water can exist on the surface (Figure \ref{fig:earth}a). Second, life must be capable of producing oxygen (Figure \ref{fig:earth}b). Here we assume Earth-like, oxygenic photosynthetic organisms (see eq. \ref{eq:photo}). Finally, the geological and atmospheric conditions must be such that the biosphere has sufficient access to nutrients such that the net biological production of oxygen is: a) greater than zero, and b) significantly greater than the geological (abiotic) production of oxygen (Figure \ref{fig:earth}c).
\subsection{Geochemical Cycles Relevant to Climate and Oxygen Production}
\subsubsection{Climate: The Carbonate-Silicate Cycle}
\begin{figure*}
	\begin{minipage}[t]{0.65\textwidth}
		\gridline{\fig{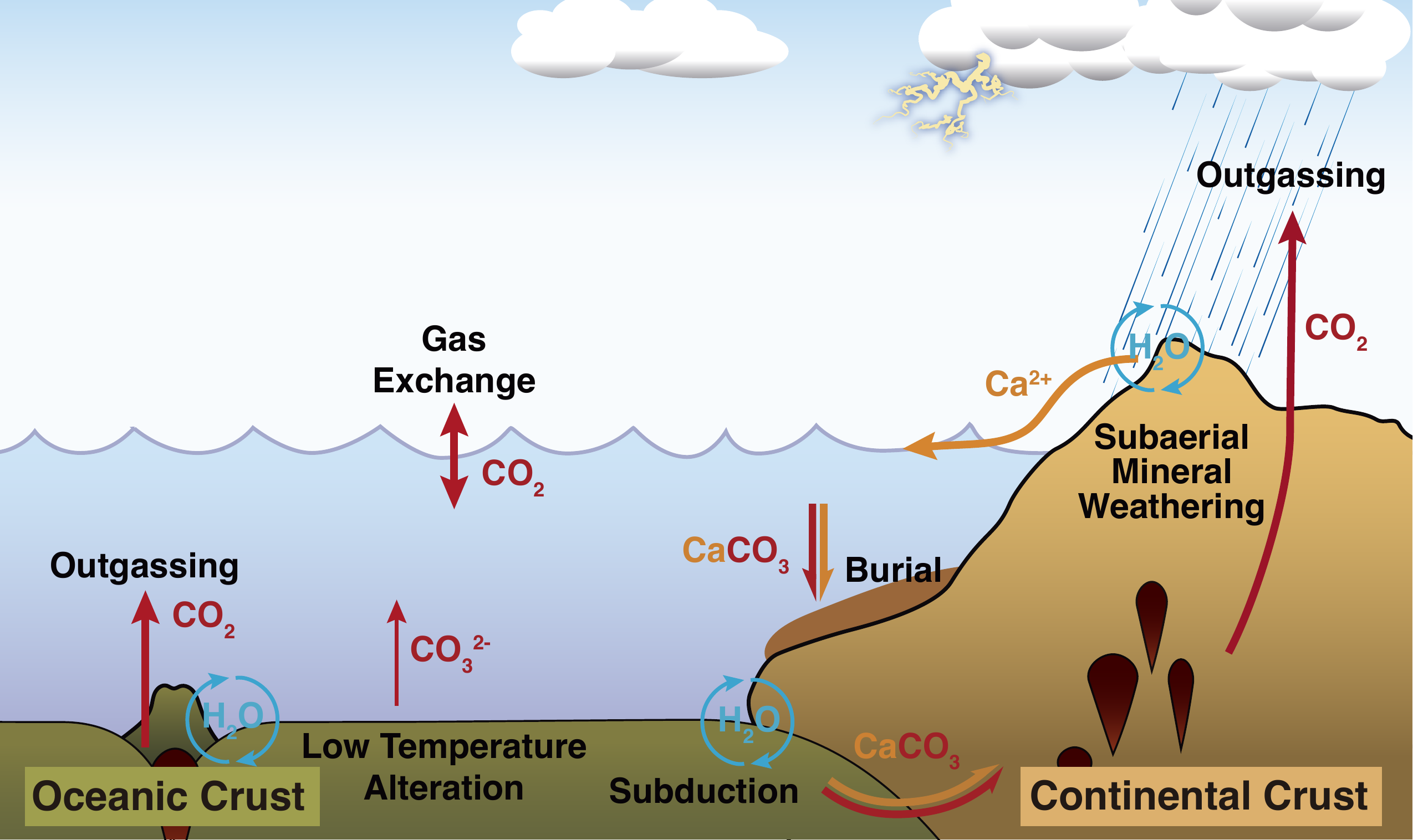}{\linewidth}{(a)}}
		\gridline{\fig{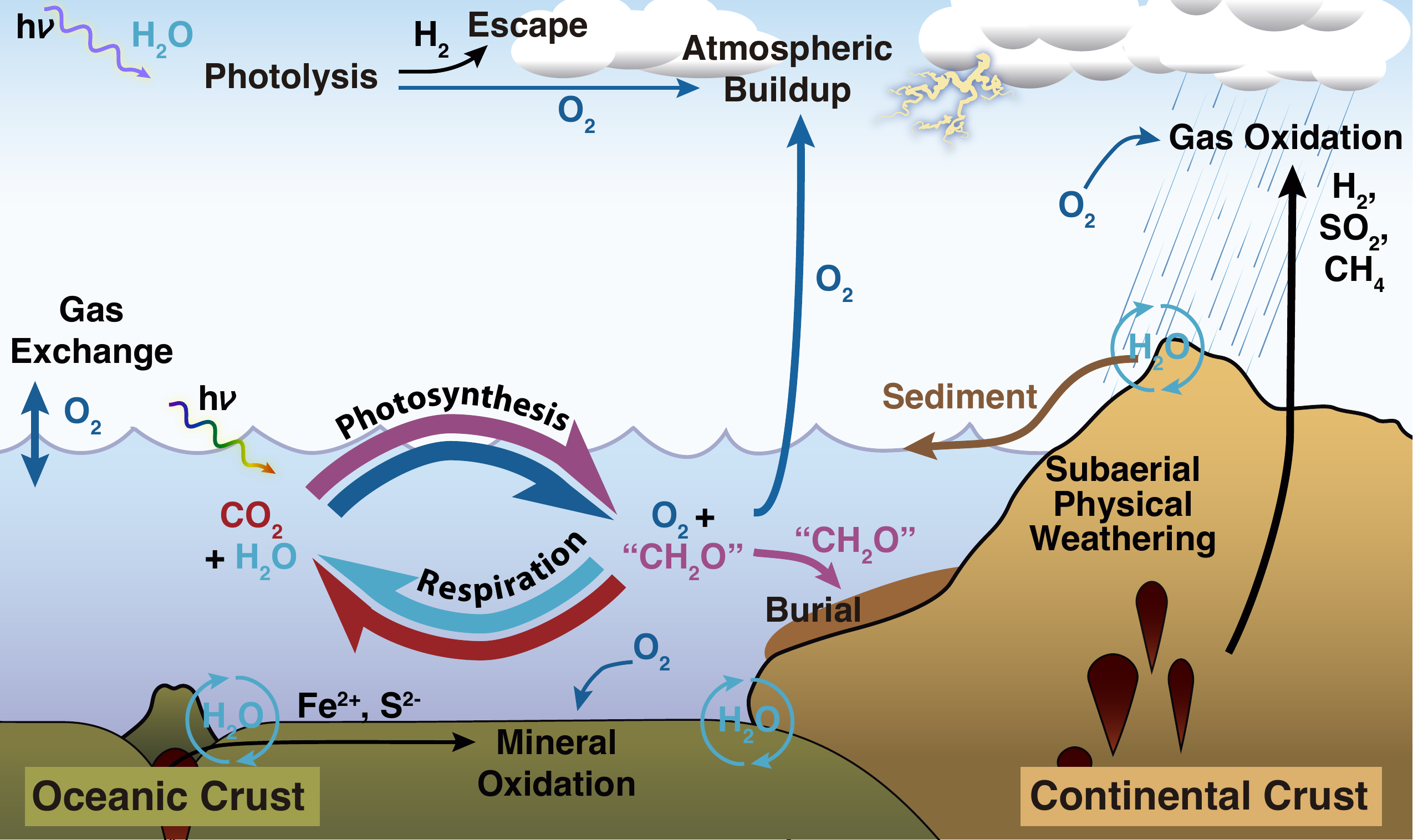}{\linewidth}{(b)}}
		\gridline{\fig{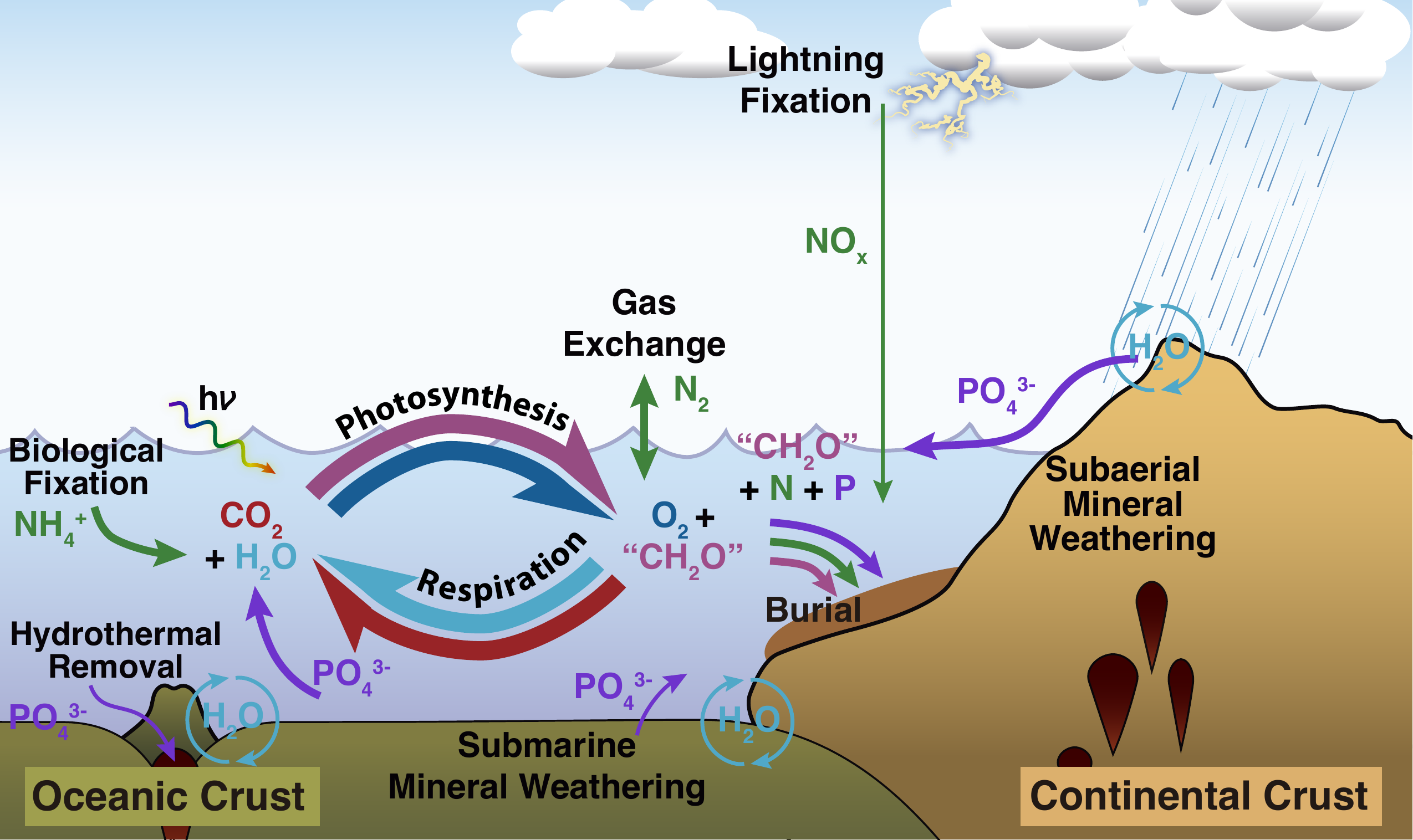}{\linewidth}{(c)}}%
	\end{minipage}
	\begin{minipage}[t]{0.32\textwidth}
		\caption{Biogeochemical cycles relevant to habitability (a), oxygen accumulation (b), and biologic oxygen production limitations (c) on Earth. Element transport and/or reactions are shown using colored lines and element labels where red is carbon dioxide (or carbonate), orange is calcium, cyan is water, magenta is organic carbon, blue is oxygen, brown is sediment, black is reductants (e.g. H\textsubscript{2} \& Fe\textsuperscript{2+}), green is nitrogen, and purple is phosphorus. Arrow widths correlate roughly with the magnitudes of the rates. Active processes have solid lines and inactive or greatly reduced processes have dashed lines. Red blobs suggest the presence of melt and/or tectonic processes. Note: outgassing and weathering can occur on both mafic and felsic crust and may have different rates for the two different environments.}
		\label{fig:earth}
	\end{minipage}
\end{figure*}

Earth’s long-term climate is stabilized by the carbonate-silicate cycle, as illustrated in figure \ref{fig:earth}a. This feedback cycle controls the CO\textsubscript{2} content of the atmosphere, and therefore modulates the acidity of rainwater and the oceans, as well as the surface temperature, through the greenhouse effect \citep{bernerPhanerozoicAtmosphericOxygen2003, bernerLongtermCarbonCycle2003, walkerNegativeFeedbackMechanism1981}. The negative feedback of the carbonate-silicate cycle may have helped Earth retain habitable surface conditions throughout its multi-billion-year history. For example, the ocean-atmosphere system contains roughly 3 x 10\textsuperscript{6} Tmol CO\textsubscript{2} which is held at steady state by a combination of production by volcanic degassing (+3.7 Tmol CO\textsubscript{2} yr\textsuperscript{-1}) and continental weathering (+22 Tmol CO\textsubscript{2} yr\textsuperscript{-1}), and consumption by carbonate deposition (–24.1 Tmol CO\textsubscript{2} yr\textsuperscript{-1}) and hydrothermal carbonatization of rocks \cite[–1.6 Tmol CO\textsubscript{2} yr\textsuperscript{-1}; e.g.,][]{sleepCarbonDioxideCycling2001}.

The key reaction, and the one that introduces climate-stabilizing negative feedback, is the weathering of continental carbonates (e.g., calcite [CaCO\textsubscript{3}]), coupled with carbonate deposition \citep{bernerLongtermCarbonCycle2003, bernerPhanerozoicAtmosphericOxygen2003, bernerWeatheringItsEffect1993}. In short, if CO\textsubscript{2} builds up in the atmosphere, the surface temperature increases through the greenhouse effect; increased temperature causes increased silicate weathering because the rate at which silicates can be exposed and physically weathered increases with increased temperature and rainfall \citep{walkerNegativeFeedbackMechanism1981}. This increased weathering rate increases the drawdown of CO\textsubscript{2}, thereby slowing the buildup of CO\textsubscript{2} in the atmosphere, leading to a negative feedback. This negative feedback maintains habitable surface temperatures that would otherwise continue to increase due to greenhouse warming.

\subsubsection{Oxygen Accumulation: The Role of Carbon Burial}
Oxygen is biologically produced on Earth by oxygenic photosynthesis using carbon dioxide (eq. \ref{eq:photo}); the reverse reaction (respiration) consumes oxygen and produces carbon dioxide. If these reactions were equally balanced through Earth’s history, the atmosphere would be anoxic. Oxygen only builds up in the Earth’s atmosphere as a result of organic carbon burial in sediments, as illustrated in figure \ref{fig:earth}b. Most of the organic carbon produced during photosynthesis is rapidly oxidized (respired) using the oxygen that is also produced during photosynthesis. It is a small imbalance in production relative to consumption that generates oxygen. This \textit{net} biological production of O\textsubscript{2} although only a fraction of the gross production rate by photosynthesis, occurs at a rate that is orders of magnitude greater than the geological oxygen production rate (due to photolysis of water followed by hydrogen escape). Burial of organic C is strongly influenced by the presence of ballast minerals, such as clays, carbonates, or silicates \citep{armstrongNewMechanisticModel2001, keilSorptivePreservationLabile1994, kennedyMineralSurfaceControl2002, mayerSurfaceAreaControl1994, ransomComparisonPelagicNepheloid1998}; organic carbon is delivered to the seafloor very efficiently when adsorbed to sinking particles. When organic carbon is buried rapidly, it can escape oxidation. Therefore, the oxygen accumulation rate on Earth is controlled both by the fine-grained sedimentation rate and by the respiration (carbon oxidation) rate. Physical weathering on land is the primary source of fine-grained sediments; it provides a control on organic carbon export from the continents \citep{galyGlobalCarbonExport2015}. Sedimentation rates on the continental margins can be as high as 200 mm kyr\textsuperscript{-1}, compared to $\sim$1 mm kyr\textsuperscript{-1} in the deepest ocean basins. As a result, on Earth, $>$ 90\% of organic C is buried on continental margins \citep{bernerBiogeochemicalCyclesCarbon1989, bernerBurialOrganicCarbon1982, burdigeBurialTerrestrialOrganic2005}. It is difficult to precisely constrain the rate at which C is buried in sediments, but it is estimated that 18.4 $\pm$ 7.8 Tmol O\textsubscript{2} yr\textsuperscript{-1}  is added to the atmosphere by these processes, all of which are biologically mediated \citep{catlingGreatOxidationEvent2014}.
\subsubsection{Limitations on Oxygen Production: Nitrogen and Phosphorus}
Earth’s atmospheric abundance of oxygen is controlled both by the rate at which it is produced and the rate at which it destroyed (i.e., reduced, removed from the atmosphere). The oxygen production rate is primarily limited by the availability of essential nutrients like N and P. The oxygen destruction rate is controlled by the abundance of geochemical reductants that are supplied by the same physical processes that weathers silicates and sediments.

Geochemical cycles relevant to nitrogen and phosphorus supply are shown in figure \ref{fig:earth}c. Nitrogen is ubiquitous in life — in the nucleotide bases of DNA and RNA, in amino acids, and in adenosine triphosphate (ATP, the energy currency of life). Similarly, P is present in DNA, RNA, and ATP as well as in cell membranes as phospholipids. In bulk plankton (phytoplankton and zooplankton), for every 106 C atoms there are 16 N atoms and 1 P atom. This 106:16:1 proportion is known as the Redfield ratio \citep{redfieldProportionsOrganicDerivatives1934}. Small variations about the canonical Redfield ratio are observed in some environments; however, the average C:N ratio in most organisms is approximately 7 and the average C:P ratio is approximately 100. On Earth, phosphorus seems to be the limiting nutrient in long-term oceanic primary production \citep{bjerrumOceanProductivityGyr2002, broeckerOcean1983, broeckerTracersSea1982, hollandChemistryAtmosphereOceans1978, tyrrellRelativeInfluencesNitrogen1999} because organisms have evolved the ability to ‘fix’ atmospheric N\textsubscript{2} into the biologically useful form, NH\textsubscript{4}\textsuperscript{+}.

Nitrogen, as N\textsubscript{2}, is the most abundant gas in Earth’s atmosphere. Despite its abundance, N often can be a limiting nutrient for life, although usually life is more limited by P. Life requires N as oxidized nitrate (NO\textsubscript{3}\textsuperscript{-}) and nitrite (NO\textsubscript{2}\textsuperscript{-}), or reduced ammonium (NH\textsubscript{4}\textsuperscript{+}). Although thermodynamically favored, reaction of N\textsubscript{2} with O\textsubscript{2} to form nitrate and nitrite is kinetically inhibited. Nitrogen fixation (conversion of N\textsubscript{2} gas into NH\textsubscript{4}\textsuperscript{+}) on modern (but preindustrial) Earth is mostly biologically mediated, at a rate of about 9 Tmol NH\textsubscript{4}\textsuperscript{+} yr\textsuperscript{-1} \citep{gruberEarthsystemPerspectiveGlobal2008}. Biological nitrogen fixation is mediated by diazotrophic microbes using nitrogenase enzymes that have weathering-derived metals (usually Mo, but occasionally Fe or V) at their reaction centers \citep{burgessMechanismMolybdenumNitrogenase1996, eadyStructureFunctionRelationships1996}. Aside from biological production, N\textsubscript{2} is also fixed by lightning strikes that cause N\textsubscript{2} and O\textsubscript{2} (or N\textsubscript{2} and CO\textsubscript{2}) to react and form NO\textsubscript{x} (nitrogen oxides) that react with water to form nitrite or nitrate \citep{franzblauElectricalDischargesInvolving1991, harmanAbioticO2Levels2018}. Fixation by lightning proceeds at a rate of $\sim$0.2 – 0.3 Tmol NO\textsubscript{3}\textsuperscript{-} yr\textsuperscript{-1} \citep{boruckiLightningEstimatesRates1984, gruberEarthsystemPerspectiveGlobal2008}. Significant atmospheric N\textsubscript{2} can be expected to persist for billions of years, and even in the absence of biological nitrogen fixation, it seems bioavailable nitrogen would be produced at the rate at which lightning can fix it, up to 0.3 Tmol NO\textsubscript{3}\textsuperscript{-} yr\textsuperscript{-1}.

The phosphorus cycle on the modern (but pre-industrial) Earth has been reviewed by Filippelli \citeyearpar{filippelliGlobalPhosphorusCycle2008}. Phosphorus is introduced into the surface environment in the form of phosphate-bearing minerals, mainly apatite minerals such as calcium fluorapatite or hydroxylapatite, e.g., [Ca\textsubscript{10}(PO\textsubscript{4})\textsubscript{6}(OH,F,Cl)\textsubscript{2}], in igneous rocks (Figure \ref{fig:earth}c). As continental rocks bearing these minerals weather and are exposed to slightly acidic rainwater, dissolution of phosphate minerals releases P-bearing ions through reactions such as the following:

\begin{equation}
\label{eq:apatite}
\scriptstyle \ce{Ca5(PO4)3OH + H+  ->[H_2O] 5Ca^2+ + 3PO4^3- + H2O}.
\end{equation}

The dissolved phosphate is transported to the oceans by rivers. Using data from Meybeck \citeyearpar{meybeckCarbonNitrogenPhosphorus1982} and Froelich et al. \citeyearpar{froelichMarinePhosphorusCycle1982}, Berner and Rao \citeyearpar{bernerPhosphorusSedimentsAmazon1994} calculated the total P flux into the oceans to be roughly 0.07 $\pm$ 0.02 Tmol P yr\textsuperscript{-1}; we adopt this same number here. The P flux to oceans is the major limit on global primary productivity, although organisms are adept at conserving and recycling P in the biosphere \citep{filippelliGlobalPhosphorusCycle2008, boyleStabilizationCoupledOxygen2014}. 

A summary of the relevant geochemical parameters for net oxygen production for Earth (and two other planet types, described below in section \ref{exo:detec}) is presented in table \ref{tab:limnut}. Additionally, oxygen is consumed by reactions with reduced gases (e.g., methane) and reduced minerals (e.g., sulfides) so that steady state is maintained. The surface environment is highly oxidizing, so the rates of these reactions are limited by the geological and biological supply of these reduced species.

\begin{deluxetable*}{lccc}
	\tablenum{1}
	\label{tab:limnut}
	\tablecaption{Summary of geochemical conditions for oxygen production on Earth, pelagic planets, and water worlds. All units are Tmol yr\textsuperscript{-1} except CO\textsubscript{2} concentration.}
	\tablewidth{0pt}
	\tablehead{\colhead{Parameter of Interest} & \colhead{Earth} & \colhead{Pelagic Planet} & \colhead{Water World}}
	\startdata
	Atmospheric CO\textsubscript{2} (bar) & $4 \times 10^{-4}$ & $10^{-4}$ to 10 & $10^{-4}$ to $>$10 \\
	Total Nitrogen Production & 9 & 0.03 & 0.03 \\
	\multicolumn{1}{r}{\textit{Biological}} & 8.7 & 0.02 & 0.02 \\
	\multicolumn{1}{r}{\textit{Non-Biological}} & 0.3 & 0.01 & 0.01 \\
	Phosphorus Production & 0.07 & $1.6 \times 10^{-4}$ & 0 to $1.6 \times 10^{-4}$ \\
	\enddata
\end{deluxetable*}

\section{Detectability of Life on Exoplanets Using Oxygen}\label{exo:detec}
\subsection{Is Life Detectable on Pelagic Planets and Water Worlds?}
To demonstrate the utility of the Detectability Index, and to illustrate our approach to using it, we apply it to putative Earth-like exoplanets. Our goal is to determine to what degree a planet can be ‘not Earth-like’ before oxygen is no longer a reliable biosignature. We specifically investigate the change in the DI for planets with different water content, and ask how much water a planet can have before DI is no longer strongly positive. For the purposes of this paper we stipulate these planets are Earth-like in all other ways, including mass, radius, bulk composition, presence of plate tectonics, a biosphere capable of oxygenic photosynthesis, etc., so that we can isolate the effects of changes in water content alone. 
We find it useful to define two planet types (Table \ref{tab:h2o}) with respect to water:

\begin{enumerate}
	\item \textquoteleft\textquoteleft pelagic planets" with 0.2 to 1 weight percent (wt\%) water at their surface, the equivalent of 8 to 40 oceans, enough to submerge all land \citep{cowanWaterCyclingOcean2014}; and
	\item \textquoteleft\textquoteleft water worlds" with $>$ 1wt\% water at their surface, an ocean that will be sufficiently deep to form high-pressure ice layers (mantles) between the liquid water and the rocky crust \citep{fuInteriorDynamicsWater2010, legerNewFamilyPlanets2004, kuchnerVolatilerichEarthMassPlanets2003}.
\end{enumerate}

We define these two classes of planets in part because water-rich planets are strongly motivated by theory and by new observations. Indeed, it is often a challenge to explain why Earth is so remarkably dry \cite[e.g.,][]{sasselovSnowLineDusty2000}. Earth’s surface hydrosphere makes up only 0.025wt\%, or 1/4000th of its total mass \citep{lecuyerHydrogenIsotopeComposition1998}. It is estimated that only a few more oceans of water are stored (technically as –OH or interstitial H) in minerals in the Earth’s mantle \citep{luthVolatilesEarthMantle2014, mottlWaterAstrobiology2007}. The Earth’s bulk water content, $\sim$0.08wt\%, is small compared to many other bodies in our own solar system, such as carbonaceous chondrites \cite[up to 13wt\% water:][]{alexanderProvenancesAsteroidsTheir2012}, Europa \cite[roughly 10wt\%:][]{andersonEuropaDifferentiatedInternal1998}, and Ceres \cite[roughly 21wt\%: ][]{deschDifferentiationCryovolcanismCharon2017, mccordCeresInternalEvolution2018}. Models in which Earth acquired its water by accreting carbonaceous chondrite-like material suggest a range of water contents from $<$ 1 ocean ($<$ 0.01wt\% H\textsubscript{2}O) to $>$ 100 oceans ($>$ 2.5wt\% H\textsubscript{2}O), with 10 – 30 oceans ($\sim$0.25 – 0.75wt\% H\textsubscript{2}O) perhaps being typical \citep{raymondMakingOtherEarths2004}. In fact, the TRAPPIST-1 system likely contains several nearly Earth-mass planets with $\geq$ 8wt\% H\textsubscript{2}O, corresponding to $\geq$ 250 Earth oceans of water \citep{gillonTemperateEarthsizedPlanets2016, unterbornInwardMigrationTRAPPIST12018}. Therefore, H\textsubscript{2}O mass fractions of up to several percent must be considered likely outcomes of planet formation.

Pelagic planets and water worlds may have geochemical cycles reasonably similar to Earth’s. Despite this similarity, increased surface water will have profound effects on geochemical processes that impact chemical alteration of rocks, ocean pH, physical weathering, sediment production, nutrient availability, and ultimately atmospheric oxygen. The existence of exposed land, we show, is a major control on the geochemical cycles, which is why we distinguish pelagic planets with no land from Earth-like planets with land. Likewise, the existence of a high-pressure ice layer between the ocean and the rocky portion of a planet can greatly impact the geochemical cycles, which is why we distinguish between pelagic planets and water worlds. Here we aim to show how increased surface water would affect each of these processes on pelagic planets and water worlds.

\subsection{Test Case 1: Pelagic Planets}
\subsubsection{Carbonate-Silicate Cycle}\label{PP:burial}
\begin{figure*}
	\begin{minipage}[t]{0.65\textwidth}
		\gridline{\fig{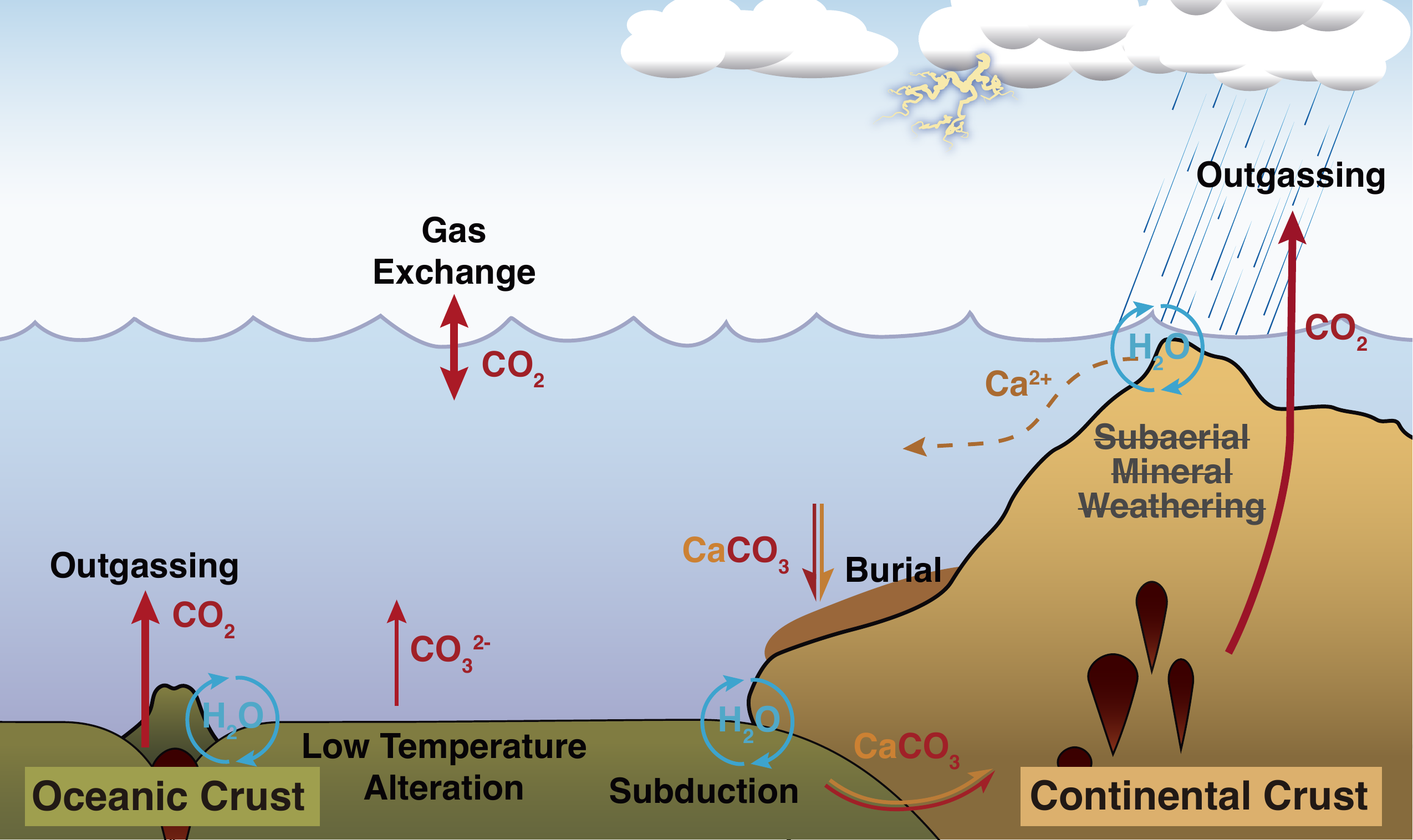}{\linewidth}{(a)}}
		\gridline{\fig{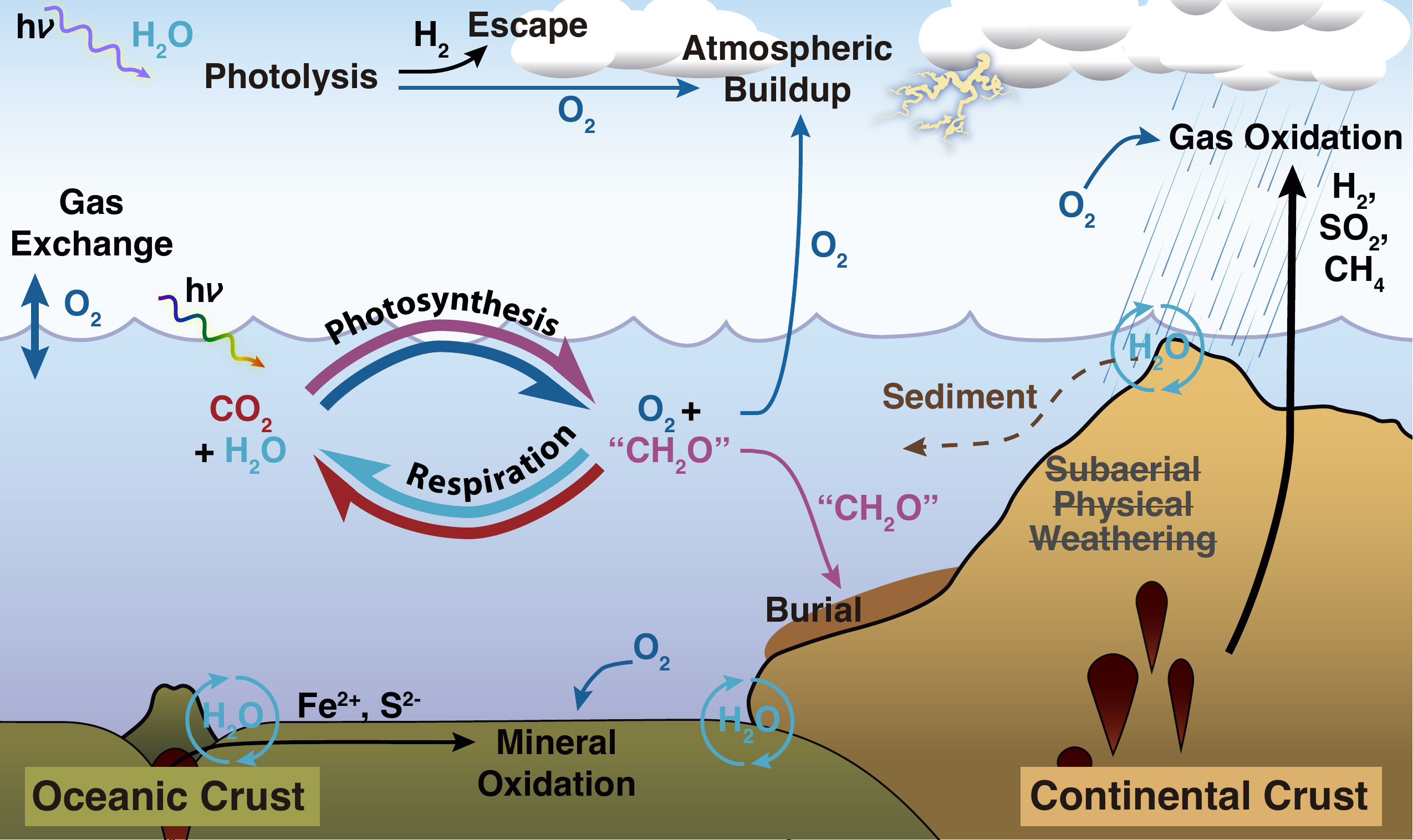}{\linewidth}{(b)}}
		\gridline{\fig{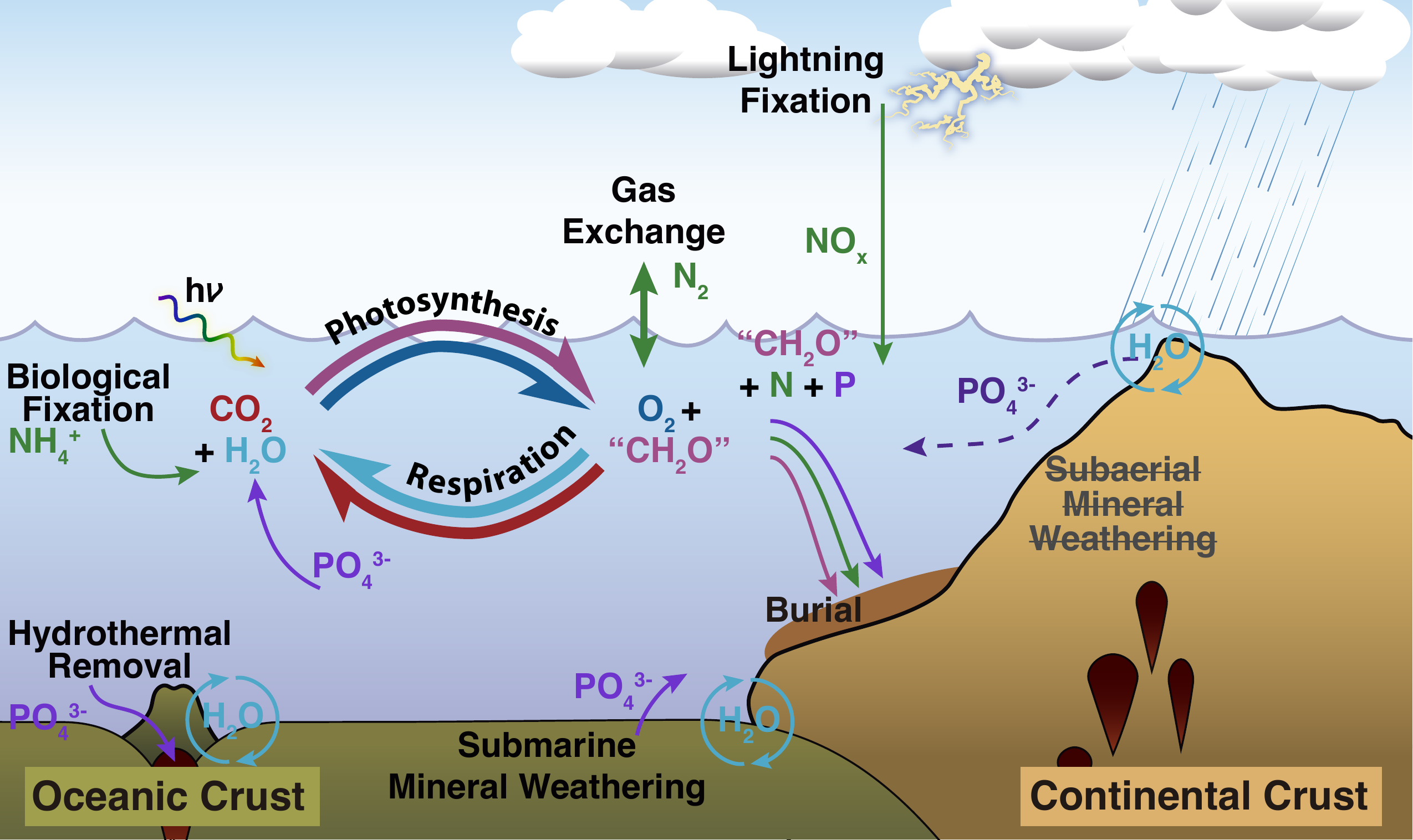}{\linewidth}{(c)}}%
	\end{minipage}
	\begin{minipage}[t]{0.32\textwidth}
		\caption{Biogeochemical cycles relevant to habitability (a), oxygen accumulation (b), and biologic oxygen production limitations (c) on pelagic planets. Element transport and/or reactions are shown using colored lines and element labels where red is carbon dioxide (or carbonate), orange is calcium, cyan is water, magenta is organic carbon, blue is oxygen, brown is sediment, black is reductants (e.g. H\textsubscript{2} \& Fe\textsuperscript{2+}), green is nitrogen, and purple is phosphorus. Arrow widths correlate roughly with the magnitudes of the rates. Active processes have solid lines and inactive or greatly reduced processes have dashed lines. Red blobs suggest the presence of melt and/or tectonic processes. Note: outgassing and weathering can occur on both mafic and felsic crust and may have different rates for the two different environments.}
		\label{fig:pelagic}
	\end{minipage}
\end{figure*}

On a pelagic planet, the carbonate-silicate cycle may act differently than on Earth, because the rate of continental weathering will be different if the continents are submerged under several kilometers of water (Figure \ref{fig:pelagic}a). We argue that mechanical weathering at the seafloor would be lower than on the continents, given that deep ocean currents have lower shear velocities than rivers, limiting their ability to entrain, expose, and mechanically weather underlying sediment. We acknowledge that seafloor turbidity currents (high-speed, destructive, albeit intermittent, gravity flows) could still occur, but their role in seafloor weathering is very poorly constrained. Many turbidity currents on Earth are initiated by river flow events from the continents and therefore, we assume there may be fewer events overall on pelagic planets. Chemical weathering will still occur, but because continental rock interacts with ocean water with relatively high pH $\approx$ 8.2, rather than with rainwater at pH $\approx$ 5, the rate may be substantially reduced \citep{liljestrandAverageRainwaterPH1985}. Another factor limiting weathering is the lowered permeability of rocks under pressure (pore spaces tend to close completely at pressures above $\sim$3 kbar \cite[e.g.,][]{walshEffectPressurePorosity1984, yangDefinitionPracticalApplication2004}. All of these factors suggest alteration of submarine rocks should be much slower than alteration of subaerial rocks. We therefore posit two end-member scenarios: one in which CO\textsubscript{2} drawdown is completely inhibited and CO\textsubscript{2} builds up in the atmosphere to tens of bars levels; and one in which CO\textsubscript{2} drawdown still proceeds, and the climate is relatively Earth-like despite the lack of continents.

A pelagic planet on which CO\textsubscript{2} built up to tens of bars would be distinctly not Earth-like. It may be difficult to know from observations of such an exoplanet exactly how much CO\textsubscript{2} was in the atmosphere, and whether temperatures at the surface would be temperate or Venus-like. It would certainly be difficult to predict the geochemical cycles with as much confidence as they could be for an Earth-like planet. Fortunately, such exoplanets also would be easily recognized by transmission spectroscopy or other observations, and planets whose atmospheres are dominated by CO\textsubscript{2} could be deprioritized.

The other outcome may be more likely, though. Sleep and Zahnle \citeyearpar{sleepCarbonDioxideCycling2001} considered the effects of removing continents from the modern carbon cycle (see their Figure 4). They asserted that chemical alteration of the newly exposed seafloor at spreading centers would still remove a comparable amount of CO\textsubscript{2}, about 3.3 Tmol CO\textsubscript{2} yr\textsuperscript{-1}. If chemical alteration proceeded at these rates, the conditions on a pelagic planet would be very Earth-like: the atmosphere would contain $\sim$1 mbar partial pressures of CO\textsubscript{2}, and the oceans would have pH near 8.

\subsubsection{Oxygen Accumulation}
Carbon burial on Earth is most efficient with high rates of sediment, delivered by physical weathering on the continents. On a pelagic planet, weathering and therefore sedimentation would be vastly different. Physical weathering of carbonates operates differently in the submarine environment compared to subaerial environments (Figure \ref{fig:pelagic}b), and is likely to lead to lower rates of carbon burial. Typical seafloor currents on Earth (driven by thermohaline circulation) flow at speeds of a few to tens of cm s\textsuperscript{-1} at most \citep{broeckerTracersSea1982}, too slow, generally, to drive significant erosion (compared to rainfall-driven erosion on Earth). It is possible that some fine-grained sediment could be generated on a pelagic planet, but the rates of geologically driven sedimentation are likely to be very much lower than on Earth. However, sedimentary particles on a pelagic planet could be derived directly from the biomass itself through two main processes. The first process is deposition of siliceous ooze, a deep-sea sediment type covering 15\% of Earth’s ocean floor, comprised primarily of the silica-bearing shells of diatoms and radiolarians. The second process involves particulate organic matter derived from heterotrophic zooplankton, which creates a flux of organic matter in open-ocean sediments of $\sim$7.6 Tmol C yr\textsuperscript{-1} on Earth. This flux derived from zooplankton is strictly governed by the rate of primary production, addressed in section \ref{PP:Lim} \citep{honjoSedimentationBiogenicMatter1982}. We can assume a 1:1 molar ratio of carbon burial to oxygen accumulation, however this assumption breaks down if organic carbon is fixed by anoxygenic photosynthesis. Considering the necessity of biologically-derived sediments for carbon burial, it is necessary to consider geochemical limitations of biological processes.

\begin{deluxetable*}{lcc}
	\tablenum{2}
	\label{tab:h2o}
	\tablecaption{Surface water mass and ocean volume for each of the example planet types.}
	\tablewidth{0pt}
	\tablehead{\colhead{Planet Type} & \colhead{Bulk H\textsubscript{2}O Mass Fraction (wt\%)} & \colhead{Oceans of Surface Water}}
	\startdata
	Earth & 0.03 & 1 \\
	Pelagic Planet & 0.2 to 1* & 8 to 40* \\
	Water World & $>$ 1\textsuperscript{$\dagger$} & $>$ 40\textsuperscript{$\dagger$} \\
	\enddata
	\tablecomments{*\citep{cowanWaterCyclingOcean2014}\\
		\textsuperscript{$\dagger$}\citep{fuInteriorDynamicsWater2010, kuchnerVolatilerichEarthMassPlanets2003, legerNewFamilyPlanets2004}}
\end{deluxetable*}

\subsubsection{Limitations to Oxygen Production}\label{PP:Lim}
On Earth, biological N fixation is co-limited by P and metals, such as Mo or Fe \citep{sanudo-wilhelmyPhosphorusLimitationNitrogen2001}. However, due to the reduced weathering rates on pelagic planets, the availability of these necessary elements may be significantly lower (figure \ref{fig:pelagic}c). Based on the P limitation reported by Sañudo-Wilhelmy et al. \citeyearpar{sanudo-wilhelmyPhosphorusLimitationNitrogen2001} and our estimated P flux (given below as $\sim$0.00016 Tmol P yr\textsuperscript{-1}), we estimate biological N fixation could be limited to $<$ 0.02 Tmol N yr\textsuperscript{-1}, because of P limitation. However, it is reasonable to assume that N\textsubscript{2} would still build up in the atmosphere of a pelagic planet because lightning could nonbiologically fix nitrogen \citep{harmanAbioticO2Levels2018}. The presence of water clouds seems sufficient to trigger lightning, as lightning is known to occur in the water clouds on Jupiter and Saturn \citep{deschProgressPlanetaryLightning2002}. The same conditions of convective thunderstorms and water clouds can be expected to exist on pelagic planets. However, lightning strikes over the ocean are much rarer than strikes over land, by a factor of about 30 \citep{williamsPhysicalOriginLand2002}. While we must consider the rate of nitrogen fixation on a pelagic planet to be highly uncertain, it is reasonable to estimate 0.3 Tmol N yr\textsuperscript{-1} $\div$ 30 $\approx$ 0.01 Tmol N yr\textsuperscript{-1}, roughly three orders of magnitude lower than biological fixation on Earth, but still potentially greater than a limited biological fixation rate on the exoplanets. This provides a floor to the production rate of biologically useful nitrogen.

The phosphorus cycle is likely to be significantly curtailed on a pelagic planet, as compared to Earth, as rates of submarine physical weathering appear to be significantly lower than rates of subaerial physical weathering (Figure \ref{fig:pelagic}c). Comparing an estimate of maximum, long-term submarine erosion rates of 35 m Myr\textsuperscript{-1} with the presumably maximum, long-term subaerial erosion rates of 200 m Myr\textsuperscript{-1}, we conclude that weathering and exposure of rock occurs at least 6 times more slowly under water than on land, albeit with much uncertainty \cite[eq.,]{duncanRadiometricAgesBasement2004, galyHigherErosionRates2001, menardRatesRegionalErosion1961, sharp50MaInitiationHawaiianEmperor2006, smootGuyotSeamountMorphology1985}. The rates of chemical dissolution that release phosphate also would be slower, due to the sensitivity of the reaction to pH as shown in the equation:

\begin{equation}
	\label{eq:Psol}
	\scriptstyle \log_{10}R_{{\rm P}} = -n{\rm H^{+}} \times p{\rm H}
\end{equation}

where R\textsubscript{P} is the rate of P release (e.g., moles of phosphate released per time, per g of rock) and nH\textsuperscript{+} is a mineral-specific constant, equal to 0.82 for fluorapatite \citep{adcockReadilyAvailablePhosphate2013, chairatKineticsMechanismNatural2007}. In brief, the reaction will proceed 77 times slower at a pH of 8 compared to the pH of rainwater at 5.7. The combined effects of reduced physical weathering and reduced chemical dissolution reduce the input rate of phosphate into the ocean by a combined factor $\sim$440 compared to Earth. We therefore consider the maximum input rate of phosphate into the oceans to be $\sim$0.00016 Tmol P yr\textsuperscript{-1}. As the phosphate supply rate from weathering is lower on a pelagic planet than on Earth, the phosphate concentrations in its oceans may also be reduced compared to Earth’s deep-sea concentration of 50 nM \citep{wheatOceanicPhosphorusImbalance2003}. Using a simple box model like that of Broecker \citeyearpar{broeckerOcean1983}, we estimate the average phosphate concentrations on a pelagic planet might be as low as 1.6 nM, approaching the limits of what the most common open ocean cyanobacterium, \textit{Prochlorococcus}, needs to reproduce.

Comparing the two nutrient fluxes, 0.03 Tmol N yr\textsuperscript{-1} and 0.00016 Tmol P yr\textsuperscript{-1}, it is likely that P will be the limiting nutrient in the biosphere, even considering that life consumes roughly 16 times as much N as P (assuming a Redfield ratio of 16:1 N:P). Considering a P-limited planet with no non-biologically produced sediments (as described in section \ref{PP:burial}), and a range of Earth-like P:C ratio in buried carbon (0.1 – 0.4\%), we identify a range of carbon burial rates 0.04 – 0.2 Tmol C yr\textsuperscript{-1}, which produces a net O\textsubscript{2} export rate to the atmosphere of 0.04 – 0.2 Tmol O\textsubscript{2} yr\textsuperscript{-1} on pelagic planets \citep{froelichMarinePhosphorusCycle1982}:

\begin{equation}
	\label{eq:Pburial}
	\begin{aligned}
		\scriptstyle 0.00016\;{\rm Tmol\;P\;yr^{-1}} \times 250\;{\rm C{:}P} \times 1\;{\rm O_2{:}C} = 0.04\;{\rm Tmol\;O_2\;yr^{-1}}\\
		\scriptstyle 0.00016\;{\rm Tmol\;P\;yr^{-1}} \times 1000\;{\rm C{:}P} \times 1\;{\rm O_2{:}C} = 0.2\;{\rm Tmol\;O_2\;yr^{-1}}
	\end{aligned}
\end{equation}

\subsection{Test Case 2: Water Worlds}
\subsubsection{Carbonate-Silicate Cycle}\label{WW:CS}
Water worlds are defined to be those with enough surface water to create pressures of at least $\sim$2 GPa at their seafloors, producing a high-pressure ice layer between the ocean and the rocky interior \citep{legerNewFamilyPlanets2004}. Many aspects of this configuration would change the carbonate-silicate cycle as we know it, creating a planet that is wholly not Earth-like (Figure \ref{fig:water}a). Even the direction of the changes is unclear. In contrast to pelagic planets, the outgassing of CO\textsubscript{2} from silicate melts very likely would be suppressed at pressures $>$ 100-200 MPa due to the presence of a high-pressure ice layer \citep{kiteHabitabilityExoplanetWaterworlds2018a}. This would effectively reduce or eliminate the influx of CO\textsubscript{2} to the atmosphere, leading to a colder planet, and one on which the carbonate-silicate cycle is no longer self-regulated \citep{cowanWaterCyclingOcean2014}. On the other hand, it may be more difficult to sequester CO\textsubscript{2} as carbonates in a deep ocean. On Earth carbonate is removed from solution through the precipitation of calcium carbonate minerals. However, if the deep ocean on a water world is undersaturated with respect to CaCO\textsubscript{3}, similar to modern Earth (i.e., below the carbonate compensation depth, $\sim$3000-5000 m in Earth's modern ocean), the long-term precipitation and deposition of calcium carbonate would be virtually impossible. This effect is due to the (atypical) increase in the solubility of calcium carbonate with increasing pressure and decreasing temperature. It is worth noting that other carbonate minerals, for example sodium carbonate, exhibit the typical decreased solubility (increased mineral stability) with increases in pressure and decreases in temperature. Again, these would be rather not Earth-like conditions. It is possible that the water world ocean could be saturated with respect to CaCO\textsubscript{3}, enabling the precipitation and burial of carbonate minerals. Therefore it is conceivable that some form of carbonate-silicate cycle might be maintained i) if water existed at the ice-rock interface \cite[cf.,][]{noackWaterrichPlanetsHow2016, kalousovaMeltingHighPressure2018, kalousovaTwophaseConvectionGanymede2018} allowing weathering of rock, and the products of weathering were advected through the ice layer, ii) if precipitation of sodium carbonate were enabled \cite[cf.,][]{marionCarbonateMineralSolubility2001}, and iii) if this precipitate were advected back down through the ice layer. Whether all these effects could occur and enable an Earth-like negative feedback against a runaway greenhouse effect, is unclear. As with pelagic planets, if CO\textsubscript{2} were not sequestered and did build up in the atmosphere of a water world sufficient to lead to very different ocean chemistry, it could be detected, and such a planet deprioritized in follow-up observations.
\subsubsection{Oxygen Accumulation}
\begin{figure*}
	\begin{minipage}[t]{0.65\textwidth}
		\gridline{\fig{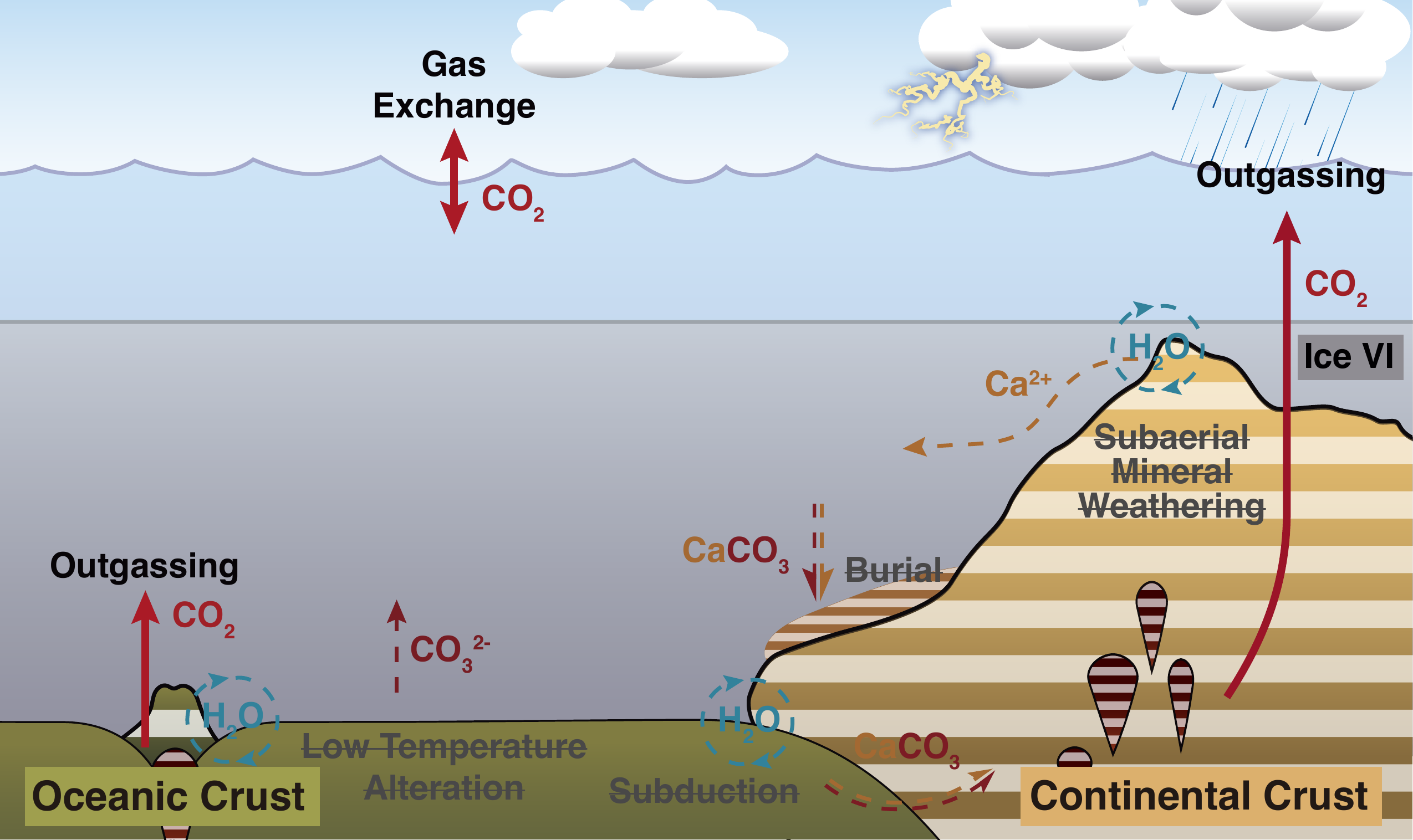}{\linewidth}{(a)}}
		\gridline{\fig{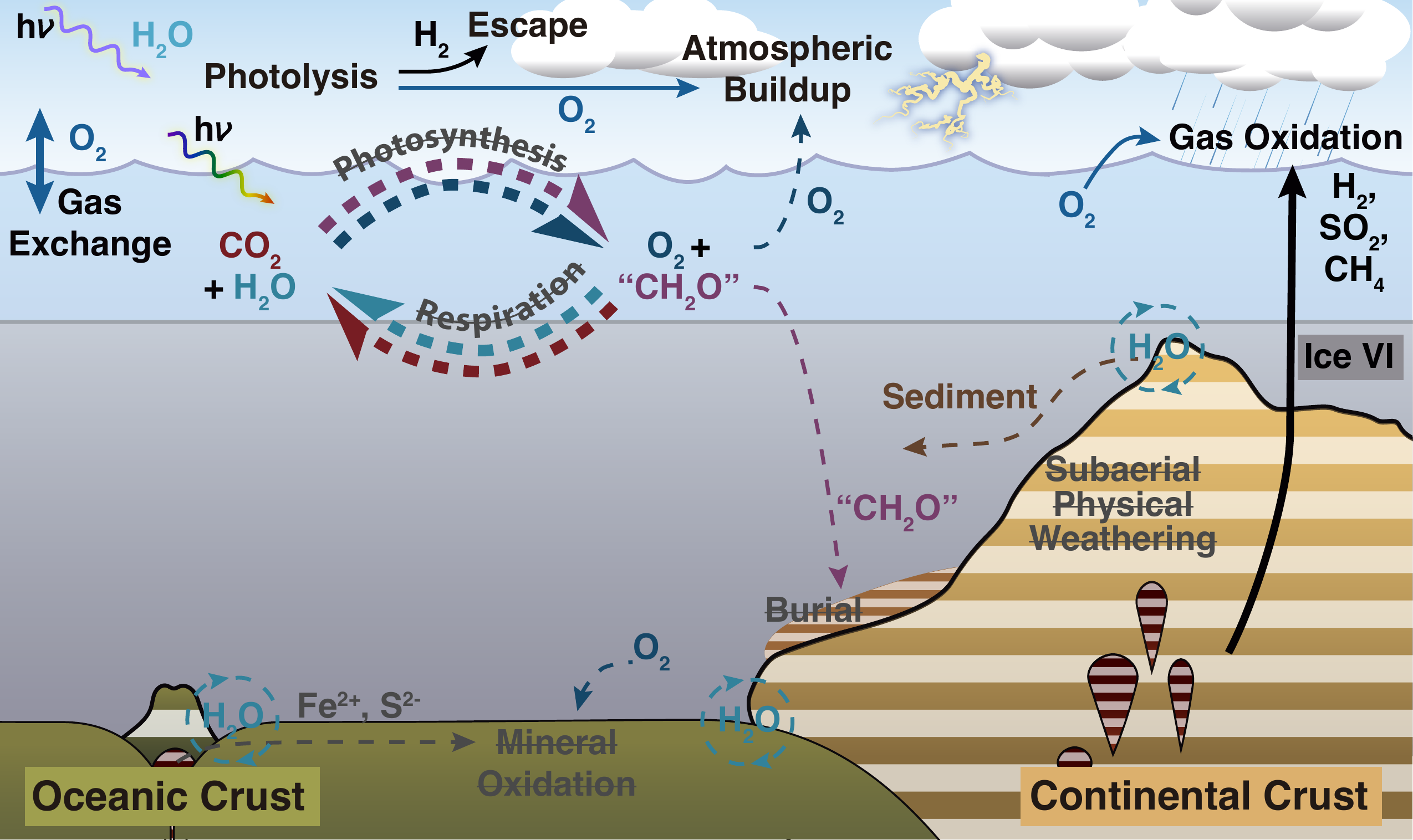}{\linewidth}{(b)}}
		\gridline{\fig{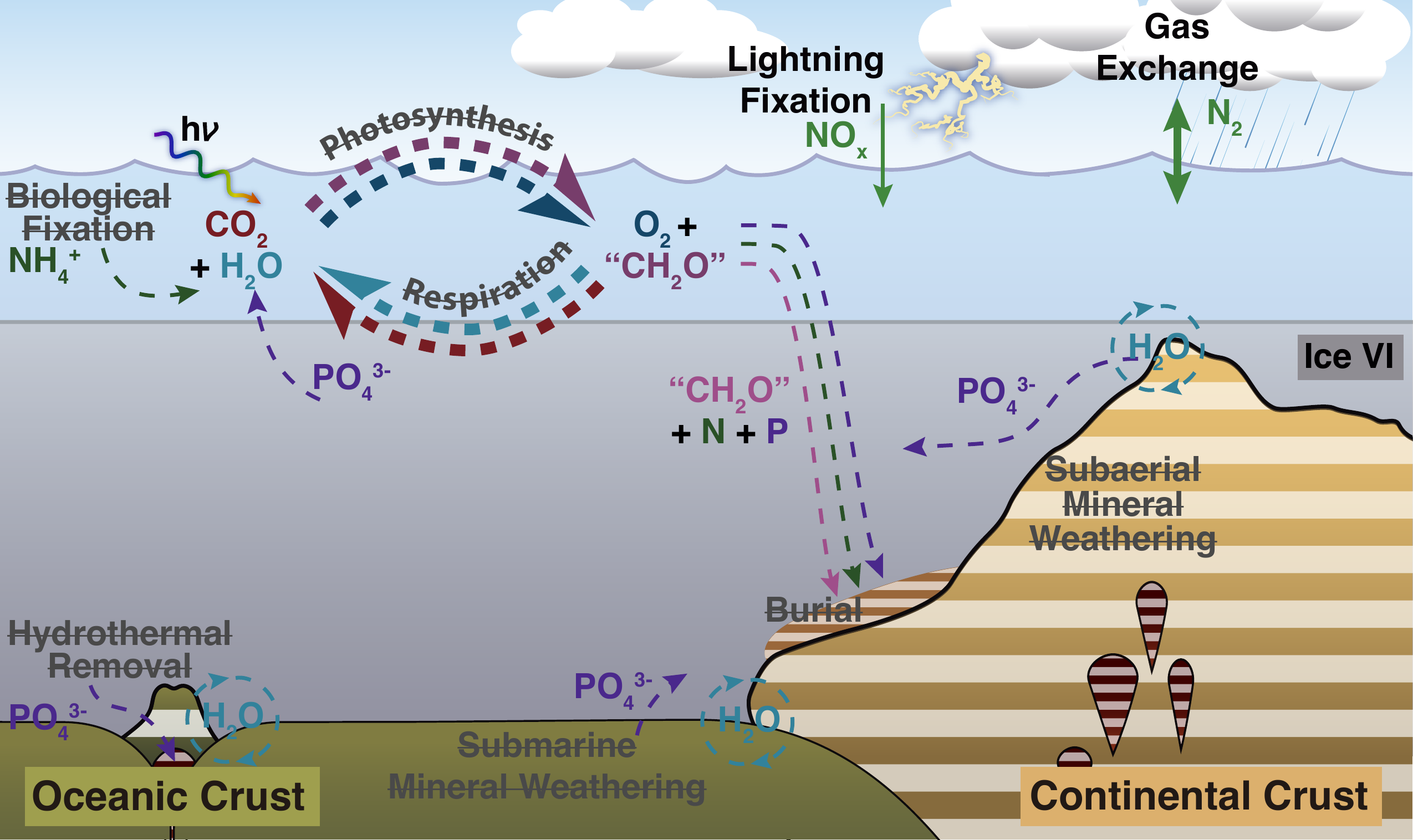}{\linewidth}{(c)}}%
	\end{minipage}
	\begin{minipage}[t]{0.32\textwidth}
		\caption{Biogeochemical cycles relevant to habitability (a), oxygen accumulation (b), and biologic oxygen production limitations (c) on water worlds. Element transport and/or reactions are shown using colored lines and element labels where red is carbon dioxide (or carbonate), orange is calcium, cyan is water, magenta is organic carbon, blue is oxygen, brown is sediment, black is reductants (e.g. H\textsubscript{2} \& Fe\textsuperscript{2+}), green is nitrogen, and purple is phosphorus. Arrow widths correlate roughly with the magnitudes of the rates. Active processes have solid lines and inactive or greatly reduced processes have dashed lines. Red blobs suggest the presence of melt and/or tectonic processes. Grey layer indicates the presence of a high pressure ice mantle. Note: outgassing and weathering can occur on both mafic and felsic crust and may have different rates for the two different environments. Grey hash marks on continental crust, melt features, hydrothermal vents, and sediment indicate the uncertainty in the presence of these features on water worlds.}
		\label{fig:water}
	\end{minipage}
\end{figure*}

Very likely the rate of organic burial on a water world is quite similar to that on a pelagic planet. However, this must be considered a maximum rate since it assumes a biological source of sedimented material, which could be severely limited by nutrient availability (see section \ref{WW:nutlim}) depending on the water world’s ice layer (Figure \ref{fig:water}b).

\subsubsection{Limitations to Oxygen Production}\label{WW:nutlim}
The nitrogen cycle on a water world would proceed very similarly to a pelagic planet, with a total flux of nonbiologically fixed nitrogen $\sim$0.01 Tmol N yr\textsuperscript{-1}. The phosphorus cycle, on the other hand, could be shut down entirely (Figure \ref{fig:water}c). In the case where there is liquid water below the ice layer, the sediment pore space will be much reduced due to increased pressure, which will lead to decreased seafloor weathering \citep{neveuCoreCrackingHydrothermal2015, walshEffectPressurePorosity1984, yangDefinitionPracticalApplication2004}. Additionally, the pH of the sub-ice layer water would likely be greater than 8 and possibly as high as 12, similar to Enceladus, greatly reducing the potential for chemical weathering \citep{gleinPHEnceladusOcean2015}. Similarly, in the case where the ice layer is directly in contact with solid rock, P production could cease almost entirely without an aqueous phase. Compounding this issue, we note that at slightly higher water fractions, pressures at the base of the high-pressure ice mantle are so high that silicate decompression melting is suppressed such that thin or perhaps no melt layers are possible \citep{kiteHabitabilityExoplanetWaterworlds2018a}. Therefore, we consider a maximum rate of net export of O\textsubscript{2} to the atmosphere of water worlds to be $\sim$0.1 Tmol O\textsubscript{2} yr\textsuperscript{-1}, limited by the lack of fine-grained sediment (see section \ref{WW:CS}) and phosphorus. However, it is more likely that Sink\textsubscript{geo} $>$ Source\textsubscript{bio} on water worlds, and that the net export of O\textsubscript{2} to the atmosphere of water worlds is $\sim$0 Tmol O\textsubscript{2} yr\textsuperscript{-1}, due to P limitation.

\subsection{Geological Oxygen Sinks: Pelagic Planets and Water Worlds}
The geological sinks for oxygen on pelagic planets and water worlds (Table \ref{tab:sum}) are estimated using results compiled here and data from Catling \citeyearpar{catlingGreatOxidationEvent2014}. Based on these estimates, it is possible that oxygen will not accumulate in these atmospheres on geological timescales, considering the oxygen sinks operate at rates an order of magnitude higher than the oxygen sources. However, it is also possible that reactions with reduced species could be limited by the supply of oxygen, and that the geological oxygen sinks may be lower. As an additional complication in the case of water worlds, if silicate melting is reduced or shut off entirely then the geological supply of reductants (as well as P-bearing minerals) will cease \citep{kiteHabitabilityExoplanetWaterworlds2018a}. In short, increased water reduces both silicate melting and magma degassing, and without melting these planets would be wholly not Earth-like and potentially geologically dead. The amount of water needed to shut off P delivery is only $\sim$2.5wt\%, equivalent to about 100 Earth oceans, but still a very small fraction compared to the typical cosmochemical abundance of water. This creates a unique case where geological production of oxygen proceeds without a geological sink, creating a false positive signal for life. In the case where Sink\textsubscript{geo} $\geq$ (Source\textsubscript{geo} + Source\textsubscript{bio}), oxygen would not be detectable in the atmosphere and the DI of the planet would be -$\infty$.

\begin{deluxetable*}{lccc}
	\tablenum{3}
	\label{tab:sum}
	\tablecaption{Geological and biological oxygen sources and sinks for Earth, pelagic planets, and water worlds. All units are Tmol yr\textsuperscript{-1}, except for the detectability index which has no unit. Estimates for pelagic planets and water worlds are based on Earth's submarine systems. All Earth values are from \citep{catlingGreatOxidationEvent2014}.}
	\tablewidth{0pt}
	\tablehead{\colhead{Reactions of Interest} & \colhead{Earth} & \colhead{Pelagic Planet} & \colhead{Water World}}
	\startdata
	\textbf{Geological (Non-biological)} & & & \\
	\textit{Oxygen Sources} & & & \\
	\multicolumn{1}{r}{Photolysis} & +0.02 & +0.02 to +0.6* & +0.02 to +0.6* \\
	\textit{Oxygen Sinks} & & & \\
	\multicolumn{1}{r}{Reduced Mineral Oxidation} & -17 & -1 & -1 to 0 \\
	\multicolumn{1}{r}{Reduced Gas Oxidation} & -5 & -4 & -4 to 0 \\
	\textit{Net Geological Oxygen Production} & -20 & -5 & -5 to 0 \\
	\textbf{Biological} & & & \\
	\textit{Oxygen Sources} & & & \\
	\multicolumn{1}{r}{Oxygenic Photosynthesis} & +10\textsuperscript{4} & $\sim$ +20 & 0 to $\sim$ +20 \\
	\textit{Oxygen Sinks} & & & \\
	\multicolumn{1}{r}{Respiration} & -10\textsuperscript{4} & $\sim$ -20 & 0 to $\sim$ -20 \\
	\textit{Net Biological Oxygen Production} & $\sim$ +20 & $\sim$ +0.04 to $\sim$ +0.2 & 0 to $\sim$ +0.2 \\
	\hline
	\textbf{Detectability Index} & +3 & -1 to +1* & -$\infty$ to +1* \\
	\enddata
	\tablecomments{* Ranges are due to differences in hydrogen escape for M stars (lower estimate) and G stars (upper estimate) }
\end{deluxetable*}

\subsection{The Importance of Co-occurring Oceans and Continents}
There are many schools of thought regarding the origin of life on Earth. In many of them, the presence of continents as well as shallow oceans is thought to be critical \citep{allwoodStromatoliteReefEarly2006, guzmanPhotoproductionLactateGlyoxylate2010, mulkidjanianOriginLifeZinc2009, mulkidjanianOriginFirstCells2012, sleepGeologicalGeochemicalConstraints2018, sugitaniBiogenicityMorphologicallyDiverse2010}. The weathering and maturation of continental crust may have created the ideal conditions for oxygenic photosynthetic organisms, and it is argued that cyanobacteria (the first oxygenic phototrophs) may have originated on land and moved to the oceans \citep{battistuzziGenomicTimescaleProkaryote2004, ponce-toledoEarlyBranchingFreshwaterCyanobacterium2017, sleepEvolutionaryEcologyRise2008}. It is intriguing that not only the detection of life by oxygen, but the existence of life that can produce oxygen, may require the co-occurrence of oceans and subaerial continents.
\section{Conclusions and Summary}
For exoplanets with different amounts of water on their surfaces, we have examined the likely biological and non-biological geochemical cycles, especially those relevant to export of O\textsubscript{2} to their atmospheres. Oxygen (as O\textsubscript{2} or O\textsubscript{3}) is a gas that potentially can be measured in the atmosphere of an exoplanet. If the rates at which reduced gases and minerals on the surface consume O\textsubscript{2} can be reasonably estimated, and steady state assumed, the production rate of O\textsubscript{2} can be inferred. Oxygen is considered a biosignature gas because on Earth its net biological production rate (oxygenic photosynthesis minus respiration), $\sim$20 Tmol O\textsubscript{2} yr\textsuperscript{-1}, is orders of magnitude greater than the production rate by non-biological processes (photolysis of H\textsubscript{2}O, followed by loss of hydrogen), 0.02 Tmol O\textsubscript{2} yr\textsuperscript{-1}. If observations of an Earth-like planet’s atmosphere implied an O\textsubscript{2} production rate $\sim$20 Tmol O\textsubscript{2} yr\textsuperscript{-1}, this could confidently be ascribed to life; an inferred production rate $\sim$0.02 Tmol O\textsubscript{2} yr\textsuperscript{-1} could not be attributed to life with any certainty. We have constructed a Detectability Index (DI) to quantify this difference (summarized in table \ref{tab:sum}); in this example for Earth, or Earth-like planets with $<$ 0.2wt\% H\textsubscript{2}O, DI(O\textsubscript{2}, Earth-like) = $log_{10}$(20/0.02) = +3. We have estimated this index for ‘pelagic planets’ with 0.2 to 1wt\% H\textsubscript{2}O, and for ‘water worlds’ with $>$1wt\% H\textsubscript{2}O. For pelagic planets, without subaerial weathering of continental rock, production of bioavailable phosphorus would be curtailed about 3 orders of magnitude, and the maximum rate at which an oxygenic photosynthesizing ecosystem could export O\textsubscript{2} to the atmosphere would be reduced accordingly by 3 orders of magnitude. Reduced rates of nitrogen fixation and sediment production are also potential limitations on net export of O\textsubscript{2}. We estimate DI(O\textsubscript{2}, pelagic planet) $\approx$ –1 to +1. For water worlds with high-pressure ice mantles, additional processes limit geochemical cycles important for net export of O\textsubscript{2} by life, and we estimate DI(O\textsubscript{2}, water worlds) $<$ +1. That is, on pelagic planets and water worlds, net biological production is at most a factor of 10 higher than non-biological production, and given uncertainties might be smaller than the non-biological production rate. We note that uncertainties in the geochemical cycles governing production of reduced gases and minerals may add uncertainty in the translation of oxygen content to oxygen production rate. Whatever total production rate is inferred on a pelagic planet or water world, with DI $<$ +1 it may not be possible to know with certainty that any of the production is due to life at all. Oxygen is not \textit{as reliable} a biosignature on pelagic planets and water worlds as it is on Earth-like planets; in fact, it may not be a reliable biosignature at all.

We conclude that for oxygen to be a reliable biosignature, a planet requires both subaerial land and water. The detectability of life, as measured by DI, changes sharply at the transition to pelagic planets, which effectively limits our scope to terrestrial planets with $<$ 0.2wt\% H\textsubscript{2}O. This is a lower water fraction than can be constrained observationally using exoplanet mass and radius, with typical precision $>$ 1wt\%. Thus, our proposed strategy for down-selecting exoplanets essentially should be to select those with as little water as possible, consistent with no measurable water. We advocate in future work (Lisse et al., submitted to ApJL) for incorporating the methodology developed here into a framework for planning observations of exoplanets.

In this paper, we have estimated the biological and geological rates, using relevant rates for Earth as a starting point. For an exoplanet of interest, we surely will have less precise knowledge of the biogeochemistry. Due to this limited information (see review of observation capabilities in \citep{fujiiExoplanetBiosignaturesObservational2018}), and stochasticity in planetary evolution models \citep{lenardicClimatetectonicCouplingVariations2016}, we may only be able to predict the properties of exoplanets statistically \citep{iyerCharacteristicTransmissionSpectrum2016, wolfgangProbabilisticMassRadius2016}. This uncertainty in geochemical cycles greatly restricts our observational strategy: in the search for life on exoplanets, O\textsubscript{2} will be a conclusive biosignature only for the most Earth-like of planets. Considering these limitations, we propose the deprioritization of planets with $\geq$ 0.2wt\% H\textsubscript{2}O, at least in searches for life on exoplanets using atmospheric oxygen. Pelagic planets and water worlds in the habitable zones of their stars could be perfectly habitable, and might even produce measurable amounts of atmospheric oxygen via oxygenic photosynthesis; but, because the production of O\textsubscript{2} could not be unambiguously attributed to life, life on these planets would be undetectable.

\acknowledgments

The results reported herein benefitted from collaborations and/or information exchange within NASA's Nexus for Exoplanet System Science (NExSS) research coordination network sponsored by NASA's Science Mission Directorate. We thank Everett Shock and members of the GEOPIG group for helpful discussion and Norm Sleep for his thoughtful and constructive review which greatly improved our paper. We gratefully acknowledge support from NExSS grant NNX15AD53G (PI Steve Desch). Nuno C Santos was supported by FCT/MCTES through national funds and by FEDER - Fundo Europeu de Desenvolvimento Regional through COMPETE2020 - Programa Operacional Competitividade e Internacionalização by these grants: UID/FIS/04434/2019; PTDC/FIS-AST/32113/2017 \& POCI-01-0145-FEDER-032113; PTDC/FIS-AST/28953/2017 \& POCI-01-0145-FEDER-028953. 

%

\vspace{5mm}




\bibliography{WaterworldsRefs_200401}{}
\bibliographystyle{aasjournal}



\end{document}